# LIMIT THEOREMS FOR THE GROVER WALK WITHOUT MEMORY

**CLEMENT AMPADU**

31 Carrolton Road
Boston, Massachusetts, 02132
USA
e-mail: drampadu@hotmail.com

**Abstract**

We consider the Grover walk as a 4-state quantum walk without memory in one dimension. The walker in our 4-state quantum walk moves to the left or right. We compute the stationary distribution of the walk, in addition, we obtain the weak limit theorem.

## I. INTRODUCTION

Recently N.Konno and T.Machida [1] introduced and studied a discrete-time 4-state quantum walk without memory associated with the Hadamard operator $H = \frac{1}{\sqrt{2}}\begin{bmatrix} 1 & 1 \\ 1 & -1 \end{bmatrix}$. Let $P(X_t = x)$ be the probability that the quantum walker, $X_t$, exists at position $x \in Z$ at time $t$ starting from the origin. The authors calculate the stationary distribution of $X_t$, and insists that the phenomenon known as "localization" occurs if $\limsup_{t \to \infty} P(X_t = 0) > 0$. By the Fourier analysis, they showed that $\frac{X_t}{t}$ converges weakly to a random variable as $t \to \infty$. The limit measure is described by both a $\delta$ – function corresponding to localization and a density function.

The aim of this paper is to study the discrete-time 4-state quantum walk without memory associated with the Grover operator $G = \begin{bmatrix} 0 & 1 \\ 1 & 0 \end{bmatrix}$. Concerning the previous work on limit theorems, in [2] the authors study an alternate to the standard two dimensional Grover walk, by replacing the requirement of a higher-dimensionality of the coin space with alternate directions, a generalization to a wider class of quantum walks is considered and a limit theorem for the alternate walk is presented. The discrete-time quantum walk on the $N-$cycle subject to decoherence on both the coin and position degrees of freedom is studied in [3]. By examining the evolution density matrix of the system, the authors derive some new conclusions about the asymptotic behavior of the system, in particular limiting behavior of the mutual information, viewed as a measure of quantum entanglement between the subsystems (coin and walker) is obtained. The long-time behaviors for discrete-time quantum walks with position measurements from the view point of weak convergence theorems is studied in [4]. In the situation that both the span of the position measurements and its number are simultaneously infinite as the final time goes to infinity, the authors have shown a crossover from ballistic spreading to diffusive spreading of the particle. Furthermore the authors introduced a new class of quantum walks, final-time dependent discrete-time quantum walks, and give the associated limit theorems. The discrete-time quantum walk on a graph with joined half lines is studied in [5]. It is shown that the quantum walk with an enlarged basis can be reduced to the walk on a half-line even if the initial state is asymmetric. Two types of limit theorems are obtained, the first describes the asymptotic behavior of the walk, which corresponds to localization. The second is a weak convergence theorem for the walk. The two-state quantum walk on the line defined by certain two $2 \times 2$ matrices is studied in [6]. One of the matrices operate the walk at only half-time. The authors present two limit theorems for the quantum walk, one concerning the stationary distribution of the walk, and the other being a convergence theorem for the walk. The time-dependent discrete-time quantum walks on the one-dimensional lattice is studied in [7]. The limit distribution of a two-period quantum walk defined by two orthogonal matrices is obtained. For the symmetric case, the distribution is determined by two orthogonal matrices. Morever, limit theorems for two special cases are presented. In [8] the simple quantum walk model is studied, the walk is obtained by quantisizing the simple random walk defined on a lattice such that particle hopping is allowed only between two nearest neighbor sites at each time step. The authors show that the novel structures of the limit distributions of pseudo-velocities in the quantum walk model can be explained by the theory of relativistic quantum mechanics by considering the connection between the solutions of the Dirac equation and the quantum walk model. In [9], the author investigates continuous-time quantum walks on star graphs. It is shown that the central limit theorem for a continuous-time quantum walk on star graphs for $N-$fold star power graph, which are invariant under the quantum component of adjacency matrix, converges to continuous-time quantum walk on complete graph with two vertices, and the probability of observing the walk tends to the uniform distribution. The discrete-time nearest-neighbor quantum walks on random environments is studied in [10]. Using the method of path counting, the authors present both quenched and annealed weak limit theorems for the quantum walk. In [11] the authors study the discrete-time quantum walk on the Cayley tree, where the walk is governed by the Grover transformation. In addition to showing that the quantum walk reduces to a quantum walk on a half line with a wall at the origin, they present two limit theorems, the first can be considered an asymptotic description of the walk in the limit corresponding to localization, whilst the second is a convergence theorem for the walk. The construction of quantum random walks with particles in an arbitrary faithful normal state, is studied in [12]. The author obtains a convergence theorem for walks demonstrating a thermalisation effect: the limit cocycle obeys a quantum stochastic differential equation without gauge terms. In [13], quantum walks driven by many coins is studied in the long time limit. For a certain class of three factors used to determine whether a particle is classical or quantum, the authors obtain limit theorems that one can use to see the transition from classical behavior to quantum. The quantum central limit theorem for continuous-time quantum walks is studied on odd graphs in [14]. It is shown that in the limit the odd graphs form a growing family of distance regular graphs. The continuous-time quantum walk on arbitrary graphs by spectral method decomposition is studied in [15]. The authors give the associated central limit theorems of the Lanczos algorithm used in the investigation of continuous-time quantum walk on arbitrary graphs of special types. In [16] the continuous-time quantum walk on a homogeneous tree in quantum probability theory is studied. The walk is defined by identifying the Hamiltonian of the system with a matrix related to the adjacency matrix of a tree. In particular, the author derives the quantum central limit theorem for the walk. In [17] the authors study the continuous-time quantum walk on the integers by introducing an $\infty \times \infty$ adjacency matrix on the integers, in particular, in contrast to the classical random walk for which the central limit theorem holds, a weak limit theorem for the continuous time quantum walk on the line is presented. For a one-parameter family of discrete-time quantum walks on the line and in the plane associated with the Hadamard walk, the author in [18] was able to derive weak convergence theorems in the long-time limit of all moments of the walker's pseudo-velocity. The quantized version of the classical correlated random walk is studied in [19]. Structural similarities between the classical random walk and its quantized version is clarified and associated limit theorems are obtained.

This paper is organized as follows. Section II treats the definition of our 4-state quantum walk which can partly be found in [1]. Section III presents the main results with proof. The first result concerns the stationary distribution of $X_t$, whilst the second is a convergence theorem for the 4-state Grover walk. In particular, by the Fourier analysis we show that $\dfrac{X_t}{t}$ converges weakly to a random variable as $t \to \infty$. The limit measure is described by both a $\delta-$ function corresponding to localization and a density function. Section IV is devoted to an open problem.

## II.  DEFINITION OF THE 4-STATE QUANTUM WALK

In this section we define the 4-state quantum walk (4QW) without memory. The state space of the

4-state quantum walk is composed of the following vectors: $|n,0\rangle$, $|n,1\rangle$, $|n,2\rangle$, $|n,3\rangle$, where $n \in Z$. For the chirality states we put $|0\rangle =^T [1\ 0\ 0\ 0]$, $|1\rangle =^T [0\ 1\ 0\ 0]$, $|2\rangle =^T [0\ 0\ 1\ 0]$, and $|3\rangle =^T [1\ 0\ 0\ 0]$, where $T$ denotes the transposed operator. We should remark that $|0\rangle$, $|1\rangle$ correspond to the left mover, and $|2\rangle$, $|3\rangle$ correspond to the right mover. In particular the states $|0\rangle$, $|1\rangle$, $|2\rangle$, $|3\rangle$ correspond to the movement "right→left", "left→left", "left→right", and "right→right", respectively. The position shift operator is given by

$$S = \sum_{n \in Z} |n-1\rangle\langle n| \otimes (|0\rangle\langle 0| + |1\rangle\langle 1|) + |n+1\rangle\langle n| \otimes (|2\rangle\langle 2| + |3\rangle\langle 3|).$$

The one-step time evolution operator is given by $S(I \otimes U)$ where $I$ is the (infinite) identity matrix and $U = \begin{bmatrix} 0 & 0 & a & c \\ b & d & 0 & 0 \\ a & c & 0 & 0 \\ 0 & 0 & b & d \end{bmatrix}$ where the nonzero entries of $U$ are complex numbers. To define the 4-state quantum walk let $|x\rangle$, $x \in C$, where $C$ is the set of complex numbers, be a (infinite-component) vector which denotes the position of the walker. Here the $x$ component of $|x\rangle$ is one, and the others are zero. Let $|\psi_t(x)\rangle \in C^4$ be the amplitude of the walker at position $x$ at time $t$. The 4-state QW at time $t$ is given by $|\psi_t\rangle = \sum_{x \in Z} |x\rangle \otimes |\psi_t(x)\rangle$. Recalling that $|0\rangle$ and $|1\rangle$ correspond to a left-mover and $|2\rangle$ and $|3\rangle$ correspond to a right mover, we can write $U = P + R$, where $P = \begin{bmatrix} 0 & 0 & a & c \\ b & d & 0 & 0 \\ 0 & 0 & 0 & 0 \\ 0 & 0 & 0 & 0 \end{bmatrix}$ and $R = \begin{bmatrix} 0 & 0 & 0 & 0 \\ 0 & 0 & 0 & 0 \\ a & c & 0 & 0 \\ 0 & 0 & b & d \end{bmatrix}$, then the evolution of the QW is determined by

$|\psi_{t+1}(t)\rangle = P|\psi_t(x+1)\rangle + R|\psi_t(x-1)\rangle$. The probability that the quantum walker $X_t$ is at position $x$ at time $t$, is given by $P(X_t = x) = \|\psi_t(x)\|^2$, where $\||x\rangle\|^2 = \langle x|x\rangle$. In order to obtain the limit theorems we introduce the Fourier Transform $|\hat{\psi}_t(k)\rangle$ of $|\psi_t(x)\rangle$ as follows:

$$|\hat{\psi}_t(k)\rangle = \sum_{x \in Z} e^{-ikx} |\psi_t(x)\rangle.$$ By the inverse Fourier transform we have

$$|\psi_t(x)\rangle = \frac{1}{2\pi} \int_{-\pi}^{\pi} e^{ikx} |\hat{\psi}_t(k)\rangle dk.$$ The time evolution of $|\hat{\psi}_t(k)\rangle$ is given by $|\hat{\psi}_{t+1}(k)\rangle = \hat{U}(k)|\hat{\psi}_t(k)\rangle$

where $\hat{U}(k) = R(k)U$ and $R(k) = \begin{bmatrix} e^{ik} & 0 & 0 & 0 \\ 0 & e^{ik} & 0 & 0 \\ 0 & 0 & e^{-ik} & 0 \\ 0 & 0 & 0 & e^{-ik} \end{bmatrix}$. The standard argument by induction on the time step gives $|\hat{\psi}_{t+1}(k)\rangle = \hat{U}(k)^t |\hat{\psi}_0(k)\rangle$. The probability that the quantum walker is at position $x$ at time $t$ is defined by $P(X_t = x) = \left\| \frac{1}{2\pi} \int_{-\pi}^{\pi} \hat{U}(k)^t |\hat{\psi}_0(k)\rangle e^{ikx} dk \right\|^2$.

### III. MAIN RESULTS

This section presents the main results, computing the stationary probability distribution of the 4-state QW and obtaining the associated weak limit theorem. First we recall the Grover walk, the Grover operator as is well known was first introduced by Moore and Russell in their study of quantum walks on the hypercube [20]. Based on Grover's diffusion, the operator has elements $a_{ij} = \frac{2}{d} - \delta_{ij}$, where $\delta_{ij} = \begin{cases} 1, i = j \\ 0, i \neq j \end{cases}$ and $d = 2^{\tilde{d}}$ where $\tilde{d}$ is the lattice dimension of the quantum walk. If $d = 4$, that is, $\tilde{d} = 2$, for example, then we get the most studied

Grover transformation, $G = \dfrac{1}{2}\begin{bmatrix} -1 & 1 & 1 & 1 \\ 1 & -1 & 1 & 1 \\ 1 & 1 & -1 & 1 \\ 1 & 1 & 1 & -1 \end{bmatrix}$. It follows from how the elements of the

Grover operator is defined that $a=0$, $b=1$, $c=1$, and $d=0$, that is in this paper,

$U = \begin{bmatrix} 0 & 0 & 0 & 1 \\ 1 & 0 & 0 & 0 \\ 0 & 1 & 0 & 0 \\ 0 & 0 & 1 & 0 \end{bmatrix}$. We should remark that the definition of $U$ and $R(k)$ implies that the matrix

$\tilde{U}(k)$ can be written as $\tilde{U}(k) = \begin{bmatrix} 0 & 0 & 0 & e^{ik} \\ e^{ik} & 0 & 0 & 0 \\ 0 & e^{-ik} & 0 & 0 \\ 0 & 0 & e^{-ik} & 0 \end{bmatrix}$. The eigenvalues $\lambda_j(k)$ ($j=1,2,3,4$) of

$\tilde{U}(k)$ can be computed as $\lambda_1 = -\lambda_4 = -1$, $\lambda_2 = -\lambda_3 = i$. Explicitly the normalized eigenvectors

$|v_j(k)\rangle$ corresponding to $\lambda_j(k)$ are given by $v_j = \begin{cases} \dfrac{1}{\sqrt{N_j(k)}} \begin{bmatrix} \lambda_j(k)e^{ik} \\ e^{2ik} \\ \lambda_j(k)e^{ik} \\ 1 \end{bmatrix}, j=1,4 \\[4ex] \dfrac{1}{\sqrt{N_j(k)}} \begin{bmatrix} -\lambda_j(k)e^{ik} \\ -e^{2ik} \\ \lambda_j(k)e^{ik} \\ 1 \end{bmatrix}, j=2,3 \end{cases}$

where the fractional term involving $N_j(k)$ ensures the normalization, and $N_j(k)$ is some constant dependent on $k$. The Fourier transform $|\hat{\Psi}_0(k)\rangle$ is expressed by $|v_j(k)\rangle$ as follows:

$|\hat{\Psi}_0(k)\rangle = \sum_{j=1}^{4} \langle v_j(k)|\hat{\Psi}_0(k)\rangle |v_j(k)\rangle$. Therefore we have

$|\hat{\Psi}_t(k)\rangle = \hat{U}(k)^t |\hat{\Psi}_0(k)\rangle = \sum_{j=1}^{4} \lambda_j(k)^t \langle v_j(k)|\hat{\Psi}_0(k)\rangle |v_j(k)\rangle$. By the inverse Fourier transform

$|\psi_t(x)\rangle = \sum_{j=1}^{4} \int_{-\pi}^{\pi} \lambda_j(k)^t \langle v_j(k)|\hat{\Psi}_0(k)\rangle |v_j(k)\rangle e^{ikx} \frac{dk}{2\pi}$. Since we can write $i = e^{\frac{\pi}{2}i}$, that is $i$ admits a

trigonometric representation, it follows from $|\psi_t(x)\rangle = \sum_{j=1}^{4} \int_{-\pi}^{\pi} \lambda_j(k)^t \langle v_j(k)|\hat{\Psi}_0(k)\rangle |v_j(k)\rangle e^{ikx} \frac{dk}{2\pi}$

and the use of the Riemann-Lebesgue lemma that we have

$|\psi_t(x)\rangle \sim \int_{-\pi}^{\pi} (-1)^t \langle v_1(k)|\hat{\Psi}_0(k)\rangle |v_1(k)\rangle e^{ikx} \frac{dk}{2\pi} + \int_{-\pi}^{\pi} \langle v_4(k)|\hat{\Psi}_0(k)\rangle |v_4(k)\rangle e^{ikx} \frac{dk}{2\pi}$, where

$g(t) \sim h(t)$ denotes $\lim_{t\to\infty} \frac{g(t)}{h(t)} = 1$. Let us take the initial state as

$|\hat{\Psi}_0(k)\rangle = |\psi_0(x)\rangle = \begin{cases} {}^T[\alpha\ \ \beta\ \ \gamma\ \ \mu], & \text{if } x = 0 \\ {}^T[0\ \ 0\ \ 0\ \ 0], & \text{if } x \neq 0 \end{cases}$, where $|\alpha|^2 + |\beta|^2 + |\gamma|^2 + |\mu|^2 = 1$ and

$|\hat{\Psi}_0(k)\rangle$ is understood when $x = 0$, then by

$|\psi_t(x)\rangle \sim \int_{-\pi}^{\pi} (-1)^t \langle v_1(k)|\hat{\Psi}_0(k)\rangle |v_1(k)\rangle e^{ikx} \frac{dk}{2\pi} + \int_{-\pi}^{\pi} \langle v_4(k)|\hat{\Psi}_0(k)\rangle |v_4(k)\rangle e^{ikx} \frac{dk}{2\pi}$, we can calculate

the following $\lim_{t\to\infty} P(X_{2t} = 0)$, $\lim_{t\to\infty} P(X_{2t+1} = 0)$, $\lim_{t\to\infty} P(X_{2t} = x)$ and $\lim_{t\to\infty} P(X_{2t+1} = x)$ from

which the stationary distribution for the 4-state Grover walk follows. Recall that

$P(X_t = x) = \||\psi_t(x)\rangle\|^2$, where $\|x\|^2 = \langle x|x\rangle$ and that $|v_1(k)\rangle = \frac{1}{e^{2ik}+1}\begin{bmatrix} -e^{ik} \\ e^{2ik} \\ -e^{ik} \\ 1 \end{bmatrix}$,

$|v_4(k)\rangle = \frac{1}{e^{2ik}+1}\begin{bmatrix} -e^{ik} \\ -e^{2ik} \\ e^{ik} \\ 1 \end{bmatrix}$, and $|\psi_0(x)\rangle = \begin{cases} {}^T[\alpha\ \ \beta\ \ \gamma\ \ \mu], & \text{if } x = 0 \\ {}^T[0\ \ 0\ \ 0\ \ 0], & \text{if } x \neq 0 \end{cases}$. We will calculate

$P(X_{2t} = 0)$ first. We notice that $P(X_{2t} = 0) = \|\psi_{2t}(0)\|^2 = \langle \psi_{2t}(0) | \psi_{2t}(0) \rangle$, where

$$|\psi_{2t}(0)\rangle \sim \int_{-\pi}^{\pi} \langle v_1(k) | \psi_0(0) \rangle |v_1(k)\rangle \frac{dk}{2\pi} + \int_{-\pi}^{\pi} \langle v_4(k) | \psi_0(0) \rangle |v_4(k)\rangle \frac{dk}{2\pi}.$$ However, we notice that

$$\int_{-\pi}^{\pi} \langle v_1(k) | \psi_0(0) \rangle |v_1(k)\rangle \frac{dk}{2\pi} = \int_{-\pi}^{\pi} \begin{bmatrix} \dfrac{\alpha - \beta e^{-ik} + \gamma - \mu e^{-ik}}{2 + e^{-2ik} + e^{2ik}} \\[2mm] \dfrac{-\alpha e^{ik} + \beta - \gamma e^{ik} + \mu e^{2ik}}{2 + e^{-2ik} + e^{2ik}} \\[2mm] \dfrac{\alpha - \beta e^{-ik} + \gamma - \mu e^{-ik}}{2 + e^{-2ik} + e^{2ik}} \\[2mm] \dfrac{-\alpha e^{-ik} + \beta e^{-2ik} - \gamma e^{-ik} + \mu}{2 + e^{-2ik} + e^{2ik}} \end{bmatrix} \frac{dk}{2\pi}$$ and

$$\int_{-\pi}^{\pi} \langle v_4(k) | \psi_0(0) \rangle |v_4(k)\rangle \frac{dk}{2\pi} = \int_{-\pi}^{\pi} \begin{bmatrix} \dfrac{\alpha + \beta e^{-ik} - \gamma - \mu e^{ik}}{2 + e^{-2ik} + e^{2ik}} \\[2mm] \dfrac{\alpha e^{ik} + \beta - \gamma e^{ik} - \mu e^{2ik}}{2 + e^{-2ik} + e^{2ik}} \\[2mm] \dfrac{-\alpha - \beta e^{-ik} + \gamma + \mu e^{ik}}{2 + e^{-2ik} + e^{2ik}} \\[2mm] \dfrac{-\alpha e^{-ik} - \beta e^{-2ik} + \gamma e^{-ik} + \mu}{2 + e^{-2ik} + e^{2ik}} \end{bmatrix} \frac{dk}{2\pi},$$ from which it follows upon

simplification that

$$|\psi_{2t}(0)\rangle \sim \int_{-\pi}^{\pi} \begin{bmatrix} \dfrac{\alpha - 2\mu\cos(k)}{1+\cos(2k)} \\ \\ \dfrac{\beta - \gamma e^{ik}}{1+\cos(2k)} \\ \\ \dfrac{-\beta e^{-ik} + \gamma + \mu i \sin k}{1+\cos(2k)} \\ \\ \dfrac{-\alpha e^{-ik} + \mu}{1+\cos(2k)} \end{bmatrix} \dfrac{dk}{2\pi} = \begin{bmatrix} \mu i \\ \\ \dfrac{i\gamma}{2} \\ \\ \dfrac{\beta i}{2} \\ \\ \dfrac{\alpha i}{2} \end{bmatrix}.$$ The computation implies that

$P(X_{2t} = 0) = \langle \psi_{2t}(0) | \psi_{2t}(0) \rangle = |\mu|^2 + \dfrac{1}{4}\left(|\gamma|^2 + |\beta|^2 + |\alpha|^2\right)$. Since this probability is independent of $t$, and is determined by the components of the initial state, we get

$\lim_{t\to\infty} P(X_{2t} = 0) = |\mu|^2 + \dfrac{1}{4}\left(|\gamma|^2 + |\beta|^2 + |\alpha|^2\right)$, a constant. Before we compute $\lim_{t\to\infty} P(X_{2t+1} = 0)$, we

should remark that if $x \neq 0$, then $|\psi_0(x)\rangle = \begin{bmatrix} 0 \\ 0 \\ 0 \\ 0 \\ 0 \end{bmatrix} =^T \begin{bmatrix} 0 & 0 & 0 & 0 \end{bmatrix}$, and so focusing on the integrands in

$|\psi_{2t}(x)\rangle \sim \int_{-\pi}^{\pi} \langle v_1(k) | \psi_0(x) \rangle |v_1(k)\rangle \dfrac{dk}{2\pi} + \int_{-\pi}^{\pi} \langle v_4(k) | \psi_0(x) \rangle |v_4(k)\rangle \dfrac{dk}{2\pi}$ or

$|\psi_{2t+1}(x)\rangle \sim -\int_{-\pi}^{\pi} \langle v_1(k) | \psi_0(x) \rangle |v_1(k)\rangle \dfrac{dk}{2\pi} + \int_{-\pi}^{\pi} \langle v_4(k) | \psi_0(x) \rangle |v_4(k)\rangle \dfrac{dk}{2\pi}$ we see that both of them give

$|\psi_{2t}(x)\rangle \sim \int_{-\pi}^{\pi} \langle v_1(k) | \psi_0(x) \rangle |v_1(k)\rangle \dfrac{dk}{2\pi} + \int_{-\pi}^{\pi} \langle v_4(k) | \psi_0(x) \rangle |v_4(k)\rangle \dfrac{dk}{2\pi} = \begin{bmatrix} 0 \\ 0 \\ 0 \\ 0 \end{bmatrix}$ and

$$|\psi_{2t+1}(x)\rangle \sim -\int_{-\pi}^{\pi}\langle v_1(k)|\psi_0(x)\rangle|v_1(k)\rangle\frac{dk}{2\pi} + \int_{-\pi}^{\pi}\langle v_4(k)|\psi_0(x)\rangle|v_4(k)\rangle\frac{dk}{2\pi} = \begin{bmatrix}0\\0\\0\\0\end{bmatrix}, \text{ thus it follows that if}$$

$x \neq 0$, then $P(X_{2t+1} = x) = P(X_{2t} = x) = \langle \psi_{2t}(x)|\psi_{2t}(x)\rangle = \langle \psi_{2t}(x)|\psi_{2t}(x)\rangle = 0$. In particular, if

$x \neq 0$, we have $\lim_{t\to\infty} P(X_{2t+1} = x) = \lim_{t\to\infty} P(X_{2t} = x) = 0$. Now we compute $\lim_{t\to\infty} P(X_{2t+1} = 0)$. We

notice in this case that that $P(X_{2t+1} = 0) = \||\psi_{2t+1}(0)\rangle\|^2 = \langle \psi_{2t+1}(0)|\psi_{2t+1}(0)\rangle$, where

$$|\psi_{2t+1}(0)\rangle \sim \int_{-\pi}^{\pi}\langle v_4(k)|\psi_0(0)\rangle|v_4(k)\rangle\frac{dk}{2\pi} - \int_{-\pi}^{\pi}\langle v_1(k)|\psi_0(0)\rangle|v_1(k)\rangle\frac{dk}{2\pi} = \int_{-\pi}^{\pi}\begin{bmatrix}\dfrac{2\beta e^{-ik}-2\gamma-\mu(e^{ik}-e^{-ik})}{2+e^{-2ik}+e^{2ik}}\\[6pt]\dfrac{2\alpha e^{ik}-2\mu e^{2ik}}{2+e^{-2ik}+e^{2ik}}\\[6pt]\dfrac{-2\alpha+\mu(e^{ik}+e^{-ik})}{2+e^{-2ik}+e^{2ik}}\\[6pt]\dfrac{-2\beta e^{-2ik}+2\gamma e^{-ik}}{2+e^{-2ik}+e^{2ik}}\end{bmatrix}\frac{dk}{2\pi}$$

Upon further simplification one has

$$|\psi_{2t+1}(0)\rangle \sim \int_{-\pi}^{\pi}\begin{bmatrix}\dfrac{\beta\cos k-\gamma-i(\beta-\mu)\sin k}{1+\cos(2k)}\\[6pt]\dfrac{\alpha\cos k-\mu\cos(2k)+i(\alpha\sin k-\mu\sin(2k))}{1+\cos(2k)}\\[6pt]\dfrac{-\alpha+\mu\cos(k)}{1+\cos(2k)}\\[6pt]\dfrac{(-\beta\cos(2k)+\gamma\cos k)+i(\beta\sin(2k)-\gamma\sin k)}{1+\cos(2k)}\end{bmatrix}\frac{dk}{2\pi} = \begin{bmatrix}-\dfrac{\beta i}{2}\\[6pt]\dfrac{-1}{2}\alpha i-\mu\\[6pt]-\dfrac{\mu i}{2}\\[6pt]\dfrac{-1}{2}\gamma i-\beta\end{bmatrix}$$

From which it follows that $P(X_{2t+1} = 0) = \langle \psi_{2t+1}(0) | \psi_{2t+1}(0) \rangle = \frac{1}{4} + \text{Im}(\overline{\alpha}\mu) + \text{Im}(\overline{\gamma}\beta) + |\mu|^2 + |\beta|^2$,

where we have used the fact that $|\alpha|^2 + |\beta|^2 + |\gamma|^2 + |\mu|^2 = 1$, $\text{Im}(z)$ is the imaginary part of the complex number $z$, and $\overline{z}$ is the complex conjugate.

So, for the 4-state Grover walk, we obtain the following stationary distribution for any initial state.

**Theorem 1:** $\lim_{t \to \infty} P(X_{2t} = 0) = |\mu|^2 + \frac{1}{4}\left(|\gamma|^2 + |\beta|^2 + |\alpha|^2\right)$,

$\lim_{t \to \infty} P(X_{2t+1} = 0) = \frac{1}{4} + \text{Im}(\overline{\alpha}\mu) + \text{Im}(\overline{\gamma}\beta) + |\mu|^2 + |\beta|^2$, and if $x \neq 0$,

$\lim_{t \to \infty} P(X_{2t+1} = x) = \lim_{t \to \infty} P(X_{2t} = x) = 0$, where $\text{Im}(z)$ is the imaginary part of the complex number $z$, and $\overline{z}$ is the complex conjugate.

We should remark from Theorem 1 that under the condition that the components of the initial states are assumed to be infinite, we see that $\delta(x) = \lim_{t \to \infty} P(X_{2t+1} = x) = \lim_{t \to \infty} P(X_{2t} = x) = \begin{cases} \infty, & \text{if } x = 0 \\ 0, & \text{if } x \neq 0 \end{cases}$, that is, the stationary distribution behaves exactly like the Dirac Delta function, and a spike is clearly seen at the origin, that is under this condition localization is seen to exist at the origin. If we drop the assumption of infiniteness on the components of the initial states, then a common criterion for the existence of localization is that $\lim_{t \to \infty} P(X_{2t+1} = 0) > 0$ or $\lim_{t \to \infty} P(X_{2t} = 0) > 0$. In particular, this begs the question of whether or not the 4-state Grover walk as defined in this paper has a $\delta$ – measure corresponding to localization. To answer this question we follow the Fourier analysis due to Grimmett et.al [21] that was partly used in the derivation of Theorem 1 to derive the weak limit measure. The starting point in the derivation is to notice that the $r$ – th moment of $X_t$ can be written as

$$E((X_t)^r) = \sum_{x \in \mathbb{Z}} x^r P(X_t = x) = \int_{-\pi}^{\pi} \langle \hat{\Psi}_t(k) | (D^r | \hat{\Psi}_t(k)) \rangle \frac{dk}{2\pi} = \int_{-\pi}^{\pi} \sum_{j=1}^{4} (t)_r \lambda_j(k)^{-r} (D\lambda_j(k))^r |\langle v_j(k) | \hat{\Psi}_0(k) \rangle|^2 \frac{dk}{2\pi} + O(t^{r-1})$$

where $D = i\left(\dfrac{d}{dk}\right)$ and $(t)_r = t(t-1) \times \cdots \times (t-r+1)$. Let $h_j(k) = \dfrac{D\lambda_j(k)}{\lambda_j(k)}$, then we obtain

$$\lim_{t \to \infty} E\left[\left(\dfrac{X_t}{t}\right)^r\right] = \int_{-\pi}^{\pi} \sum_{j=1}^{4} h_j^r(k) \left|\langle v_j(k) | \widehat{\Psi}_0(k) \rangle\right|^2 \dfrac{dk}{2\pi} = 0^r \Delta + \int_{-\pi}^{\pi} \sum_{\substack{j=1 \\ j \neq 1,4}}^{4} h_j^r(k) \left|\langle v_j(k) | \widehat{\Psi}_0(k) \rangle\right|^2 \dfrac{dk}{2\pi}$$

where

$$\Delta = \int_{-\pi}^{\pi} \sum_{\substack{j=1 \\ j \neq 2,3}}^{4} \left|\langle v_j(k) | \widehat{\Psi}_0(k) \rangle\right|^2 \dfrac{dk}{2\pi}$$

$$= \int_{-\pi}^{\pi} \left|\dfrac{-\alpha e^{-ik}}{e^{-2ik}+1} + \dfrac{\beta e^{-2ik}}{e^{-2ik}+1} - \dfrac{\gamma e^{-ik}}{e^{-2ik}+1} + \dfrac{\mu}{e^{-2ik}+1}\right|^2 \dfrac{dk}{2\pi} + \int_{-\pi}^{\pi} \left|\dfrac{-\alpha e^{-ik}}{e^{-2ik}+1} - \dfrac{\beta e^{-2ik}}{e^{-2ik}+1} + \dfrac{\gamma e^{-ik}}{e^{-2ik}+1} + \dfrac{\mu}{e^{-2ik}+1}\right|^2 \dfrac{dk}{2\pi}$$

and $\int_{-\pi}^{\pi} \left|\dfrac{-\alpha e^{-ik}}{e^{-2ik}+1} + \dfrac{\beta e^{-2ik}}{e^{-2ik}+1} - \dfrac{\gamma e^{-ik}}{e^{-2ik}+1} + \dfrac{\mu}{e^{-2ik}+1}\right|^2 \dfrac{dk}{2\pi} = \beta\mu + \dfrac{1}{2}[\alpha\beta + \beta\gamma + \alpha\mu + \gamma\mu]i$, and

$\int_{-\pi}^{\pi} \left|\dfrac{-\alpha e^{-ik}}{e^{-2ik}+1} - \dfrac{\beta e^{-2ik}}{e^{-2ik}+1} + \dfrac{\gamma e^{-ik}}{e^{-2ik}+1} + \dfrac{\mu}{e^{-2ik}+1}\right|^2 \dfrac{dk}{2\pi} = -\beta\mu - \dfrac{1}{2}[\alpha\beta + \beta\gamma - \alpha\mu + \gamma\mu]i$, and thus

$\Delta = \dfrac{1}{2}\alpha\mu i$. Now we let $f_K(x) = \dfrac{1}{\pi(1-x^2)\sqrt{1-2x^2}} I_{(0,1)}(x)$, where $I_A(x) = \begin{cases} 1, & \text{if } x \in A \\ 0, & \text{if } x \notin A \end{cases}$, and

$h_j(k) = x$, then we see that

$$\lim_{t \to \infty} E\left[\left(\dfrac{X_t}{t}\right)^r\right] = 0^r \Delta + \int_{-\infty}^{\infty} x^r (c_0 + c_1 x + c_2 x^2) f_K(x) dx = \int_{-\infty}^{\infty} x^r \{\Delta \delta_0(x) + (c_0 + c_1 x + c_2 x^2) f_K(x)\} dx,$$

where $c_0$, $c_1$, and $c_2$ are constants determined by the initial state $|\psi_0(0)\rangle = {}^T[\alpha \ \beta \ \gamma \ \mu]$. It should be noted that the determination of $\Delta$ as was calculated explicitly above is also determined by the initial state $|\psi_0(0)\rangle = {}^T[\alpha \ \beta \ \gamma \ \mu]$, due to the tedious computations we shall not give the general analytic formula in the case of $c_0$, $c_1$, and $c_2$.

So for the four-state Grover walk our weak limit measure which has a $\delta$ – measure corresponding to localization is given by the following.

**Theorem 2:** For $-\infty < a \leq b < \infty$, $\lim_{t \to \infty} P\left(a \leq \frac{X_t}{t} \leq b\right) = \int_a^b \Delta \delta_0(x) + \left(c_0 + c_1 x + c_2 x^2\right) f_K(x) \, dx$, where

$$f_K(x) = \frac{1}{\pi(1-x^2)\sqrt{1-2x^2}} I_{(0,1)}(x),$$ $\delta_0(x)$ denotes the Dirac's Delta function at the origin, $\Delta = \frac{1}{2}\mu i$,

and $c_0$, $c_1$, $c_2$ are constants determined by the initial state $|\psi_0(0)\rangle = ^T[\alpha \ \beta \ \gamma \ \mu]$.

## IV. Open Problem

Consider the quantum walk on the $k-$dimensional lattice governed by the $2^k \times 2^k$ unitary matrix

$U = A^{\otimes k}$, where $A = \begin{bmatrix} 0 & 0 & a & c \\ b & d & 0 & 0 \\ a & c & 0 & 0 \\ 0 & 0 & b & d \end{bmatrix}$. It is an open problem to obtain the limit theorems for the quantum walk for a general $a, b, c, d \in C$ and $m-$state. We should remark that the case $k = 1$ was proposed by Konno and Machida [1], and we believe it is still unsolved. Therefore, this problem is extreme!